\documentclass[mathleft]{an}
\usepackage{graphicx}
\begin{document}

\Pagespan{1}{}
\Yearpublication{2011}%
\Yearsubmission{2010}%
\Month{11}%
\Volume{999}%
\Issue{88}%

\title{Energy oscillations and a possible route to chaos in a 
modified Riga dynamo}

\author{F. Stefani \inst{1}\fnmsep\thanks{Corresponding author:
  \email{F.Stefani@fzd.de}\newline}
\and  A. Gailitis\inst{2}
\and G. Gerbeth\inst{1}
}
\titlerunning{Energy oscillations in a modified Riga dynamo}
\authorrunning{F. Stefani, A. Gailitis, \& G. Gerbeth}
\institute{
Forschungszentrum Dresden-Rossendorf, P.O. Box 510119, D-01314 Dresden, Germany
\and 
Institute of Physics, Salaspils, Miera iela, Latvia
}

\received{31 August 2010}
\accepted{11 Nov 2010}
\publonline{later}


\abstract{Starting from the present version of the Riga dynamo
  experiment with its rotating magnetic eigenfield dominated by a single
  frequency we ask for those modifications of this set-up that 
  would allow for a non-trivial magnetic field behaviour in the 
  saturation regime.
  Assuming an increased ratio of azimuthal
  to axial  flow velocity, we obtain energy oscillations 
  with a frequency below the eigenfrequency of the 
  magnetic field. 
  These new oscillations are identified as magneto-inertial waves
  that result from a slight imbalance of Lorentz and inertial
  forces. Increasing the azimuthal velocity further, or 
  increasing the total magnetic 
  Reynolds number, we find transitions to a
   chaotic behaviour of the dynamo.}

\maketitle

\section{Introduction}

The last decade has seen great advantages
in the experimental realization of hydromagnetic
dynamos (Stefani, Gailitis and Gerbeth 2008), culminating 
perhaps in the recent observation of reversing and 
chaotic dynamos in the French VKS experiment (Ravelet 2008).

The Riga dynamo experiment traces back to one of the 
simplest homogeneous dynamo concepts that had been proposed by 
Ponomarenko (1973).
The Ponomeranko dynamo consists of a conducting rigid rod
that spirals through  a medium of the 
same conductivity that extends infinitely in radial and axial direction.

By a detailed analysis of this dynamo,
Gailitis and Freibergs  (1976) were able to identify the
rather low  critical magnetic Reynolds number of 17.7 for the 
convective instability. 
An essential step towards the later experimental realization
was the addition of  a straight back-flow concentric to 
the inner helical flow, which converts  the 
convective instability into an absolute one 
(Gailitis and Freibergs 1980).

After many years of optimization, design and
and construction 
(cf. Gailitis et al. 2008 for a recent survey)
a first experimental 
campaign took place in November 1999, resulting in
the observation of a slowly growing magnetic eigenfield
(Gailitis et al. 2000). 

The four dynamo runs of July 2000 
provided a first  stock of 
growth rate, frequency, 
and spatial structure data of the magnetic eigenfield, 
this time both in the
kinematic as well as in the saturated regime (Gailitis et al. 2001).
During the following experimental campaigns, 
plenty of growth rate and frequency 
measurements were carried out. When scaled by the
temperature dependent conductivity of sodium, both quantities 
turned out to be reproducible over the years. 
(Note, however, that after the replacement of the outworn gliding 
ring seal by a modern magnetic coupler in 2007 a problem with 
a reduced azimuthal velocity occurred 
which is detrimental for dynamo action and which 
needs further inspection).

\begin{figure}
\includegraphics[width=80mm]{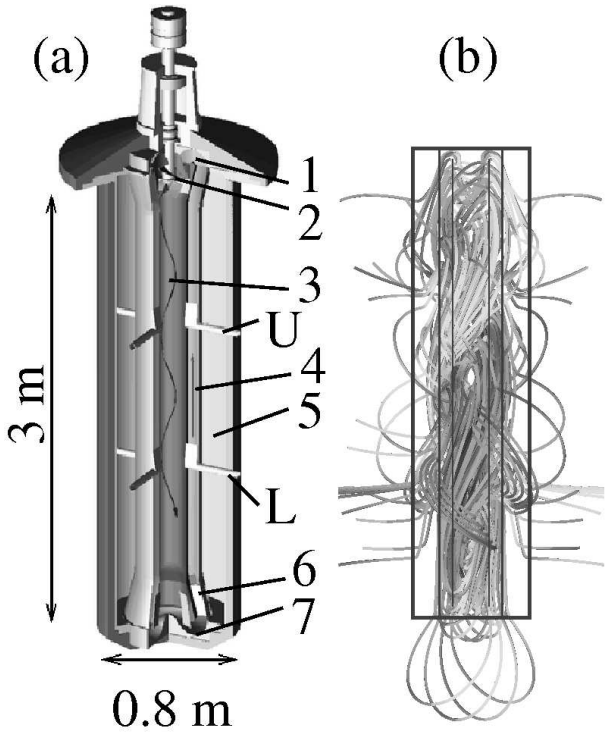}
\caption{
The Riga dynamo experiment and its eigenfield.
(a) Details of the central dynamo module. 1 - Upper bending region; 2 - Propeller;
3 - Central helical flow region; 4 - Return-flow region; 5 - Outer sodium
region; 6 - Guiding vanes for straightening the flow in the return flow. 
7 - Lower bending region. At approximately 1/3 (L) and 2/3 (U) of the 
dynamo height there are
four ports for various magnetic field, pressure and velocity probes.
(b) Simulated structure of the magnetic eigenfield                              
in the kinematic regime.
}
\label{label1}
\end{figure}

With view on the future utilization of the Riga dynamo experiment,
here we deal  with the general question whether
the facility could be modified in such a way that
it exhibits a more complicated saturation 
behaviour. 
One idea that comes to mind is the principle possibility
to lower the 
critical $Rm$ by means of a parametric 
resonance, or swing excitation,
(Rohde, R\"udiger and Elstner 1999)
with a periodic flow forcing with approximately the double of 
eigenfrequency. 
We feel, however, that the comparably high frequency 
(around 4 Hz and larger) 
that would be needed for this purpose is far beyond the 
technical specifications of the utilized 
motors and their control.

\begin{figure}
\includegraphics[width=80mm]{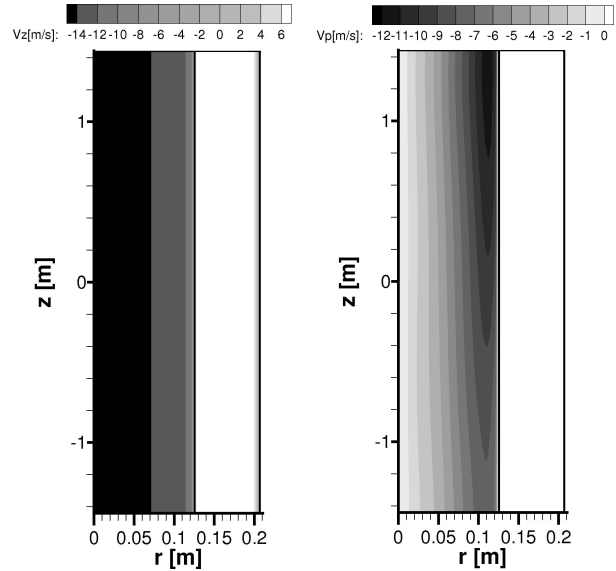}
\caption{Assumed kinematic velocity structure in the central
part of the inner tube and in the back-flow tube at
a rotation rate of 2100 rpm. Axial velocity $v_z(r,z)$ (l.h.s) and
azimuthal velocity $v_{\varphi}(r,z)$ (r.h.s.). 
Note the slight downward decay of $v_{\varphi}(r,z)$. Figure 
adopted from (Gailitis et al. 2002)}.
\label{label2}
\end{figure}

Another idea, namely to modify the present ratio of
azimuthal velocity $v_{\varphi}$ and axial
velocity $v_z$,  is motivated by the following considerations:
The Riga dynamo had been constructed 
in such a way that the flow helicity is maximum for a given 
kinetic energy (Gailitis et al. 2004) 
which implies the ratio of $v_{\varphi}$ to 
$v_z$ to be close to 1. Now, since 
the saturation mechanism relies strongly on a 
selective 
braking of $v_{\varphi}$
one could ask for the consequences 
if one would 
start with a too strong $v_{\varphi}$. 
Its braking would then lead to a 
reduced ratio $v_{\varphi}$ to $v_z$
which  might be suspected to be even 
better suited for 
dynamo action than the original 
kinematic flow. In the extreme form this mechanism 
could lead to a sub-critical Hopf bifurcation, meaning that 
the velocity in the saturated state
has a lower critical $Rm$ than the kinematic (unperturbed) velocity.
This  effect should be 
detectable in form of a hysteretic behaviour of self-excitation,
similar to the case studied recently by Reuter et al. (2008). 
But even if an "honest"
sub-critical Hopf bifurcation could not be achieved, a 
general tendency 
of the system to oscillate between a weak-field and 
a strong-field state is still to be expected.
Since such an oscillation would appear from the destabilization
of the (former) steady equilibrium 
between Lorentz forces and inertial forces,
it should be called a magneto-inertial 
wave, similar to the waves that have been observed in the
DTS-experiment in Grenoble (Schmitt 2010).

Further, if such an oscillation with a second frequency sets in, 
one could also ask for the appearance of more frequencies 
and possibly for a transition to chaos according to the 
Ruelle-Takens-Newhouse scenario (Eckmann 1981). 

These foregoing ideas delineate 
the scope of the present paper. In the second section, we 
will describe briefly the Riga dynamo, and present the 
utilized scheme for its numerical simulation. 
Then we will discuss two 
sorts of parameter variations that both 
start from the well-known 
single-frequency saturation state and end up with 
a chaotic regime.
The paper concludes with a few remarks concerning the technical 
feasibility of the proposed modification.

\section{The Riga dynamo and the numerical code for its simulation}

An illustration of the Riga dynamo, and of its magnetic
eigenfield is
shown in Fig. 1. The  facility consists basically
of three concentric cylinders. In the innermost cylinder, the
liquid sodium undergoes a downward spiral motion with 
optimized helicity. In the second cylinder, it flows back to
the propeller region. The third cylinder serves
for improving the electromagnetic boundary conditions which helps
in achieving a minimum critical $Rm$.
More details about the facility and the experimental campaigns 
can be found in (Gailitis et al. 2000, 2001, 
2003, 2004, 2008). In the present configuration, self-excitation 
occurs approximately at a propeller 
rotation rate of $P=1840$ rpm when scaled appropriately to a 
reference temperature of 157$^{\circ}$C. The shape of the 
magnetic eigenfield 
resembles a double-helix that rotates around the vertical axis with 
a frequency in the range of  1-2 Hz.

\begin{figure*}
\includegraphics[width=165mm]{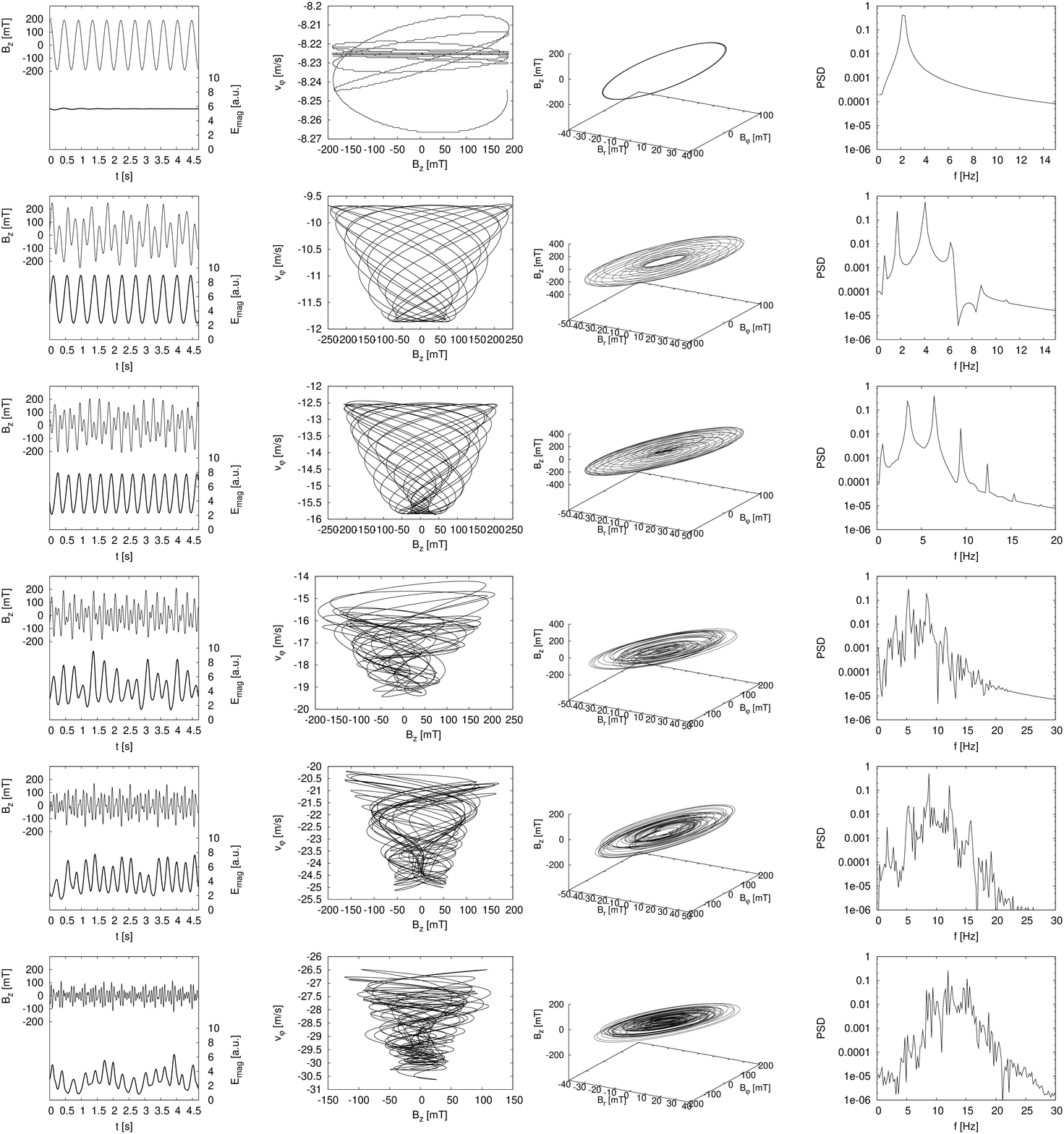}
\caption{Influence of the increasing ratio of $v_{\varphi}$ and $v_z$ 
(factor $A$ compared with the present flow structure) on the 
time evolution. From top to bottom: A=1, 1.3, 1.7, 2.0, 2.5, 3.0. From
left to right: Magnetic field energy and 
$B_z(r=0.086 \mbox{ m},z=-0.39 \mbox{ m})$ (first column), 
two-dimensional phase portrait of $B_z$  and $v_{\varphi}$ at 
$(r=0.086 \mbox{ m},z=-0.39 \mbox{ m})$ (second column), 
three-dimensional 
phase  portrait of  $B_r$, $B_{\varphi}$, and $B_z$
$(r=0.086 \mbox{ m},z=-0.39 \mbox{ m})$ (third column), 
power spectral density
of $B_z(r=0.086 \mbox{ m},z=-0.39 \mbox{ m})$ (fourth column).}
\label{label3}
\end{figure*}

The numerical simulations of this paper will be carried out by
means of 
a code that couples a solver for the induction equation
\begin{eqnarray}
\frac{\partial {{\bf{B}}}}{\partial t}=\nabla
\times ({\bf{v}} \times {\bf{B}})
+\frac{1}{\mu_0 \sigma} \Delta {\bf{B}} \label{4} \; .
\label{indeq}
\end{eqnarray}
for the azimuthal $m=1$ component of the magnetic 
field $\bf B$, and a simplified solver for the 
velocity perturbation. 
The solver for the induction equation 
is a two-dimensional finite difference scheme with a 
non-homogeneous grid in $r$ and $z$,
that uses 
an Adams-Bashforth method of second order for the time integration
(cf. Gailitis et al. 2004 for more details).
The real geometry of the dynamo module has been slightly
simplified, with all curved parts in the bending regions
being replaced by rectangular ones. 
The unperturbed velocity in the central channel
was inferred from a number of measurements at a water test facility 
and some extrapolations. These measurements had revealed a
slight downstream decay of $v_{\varphi}$.
The velocity in the back-flow channel has been 
considered as purely axial and
constant in $r$ and $z$, with the constraint 
that the volume flux there is   
equal to the volume flux in the inner cylinder. 
Figure 2 shows the assumed kinematic velocity 
components $v_z(r,z)$ and $v_{\varphi}(r,z)$
both in the central channel and in the back-flow 
channel.

\begin{figure}
\includegraphics[width=80mm]{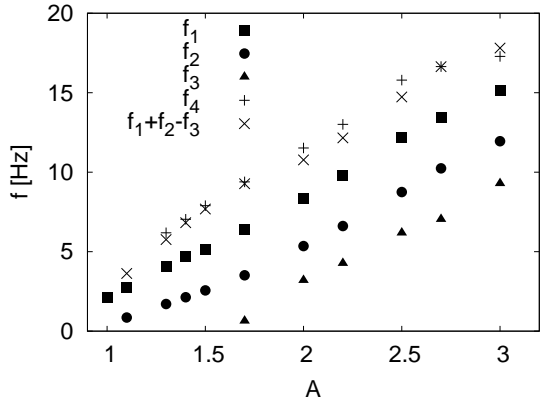}
\caption{The three lowest dominant frequencies for increasing factor $A$.
$f_1$ denotes the frequency of the magnetic eigenfield, $f_2$ 
is the frequency of the first magneto-inertial oscillation, $f_3$ 
is the third arising frequency. $f_4$ corresponds approximately to
the combination $f_1+f_2-f_3$.}
\label{label4}
\end{figure}

The induction equation solver
has been extensively used in the optimization 
of the Riga dynamo experiment, and its main results 
(in particular
concerning the growthrate and the frequency of the eigenfield)
agree to a very few percent with the experimental results. 

\begin{figure*}
\includegraphics[width=165mm]{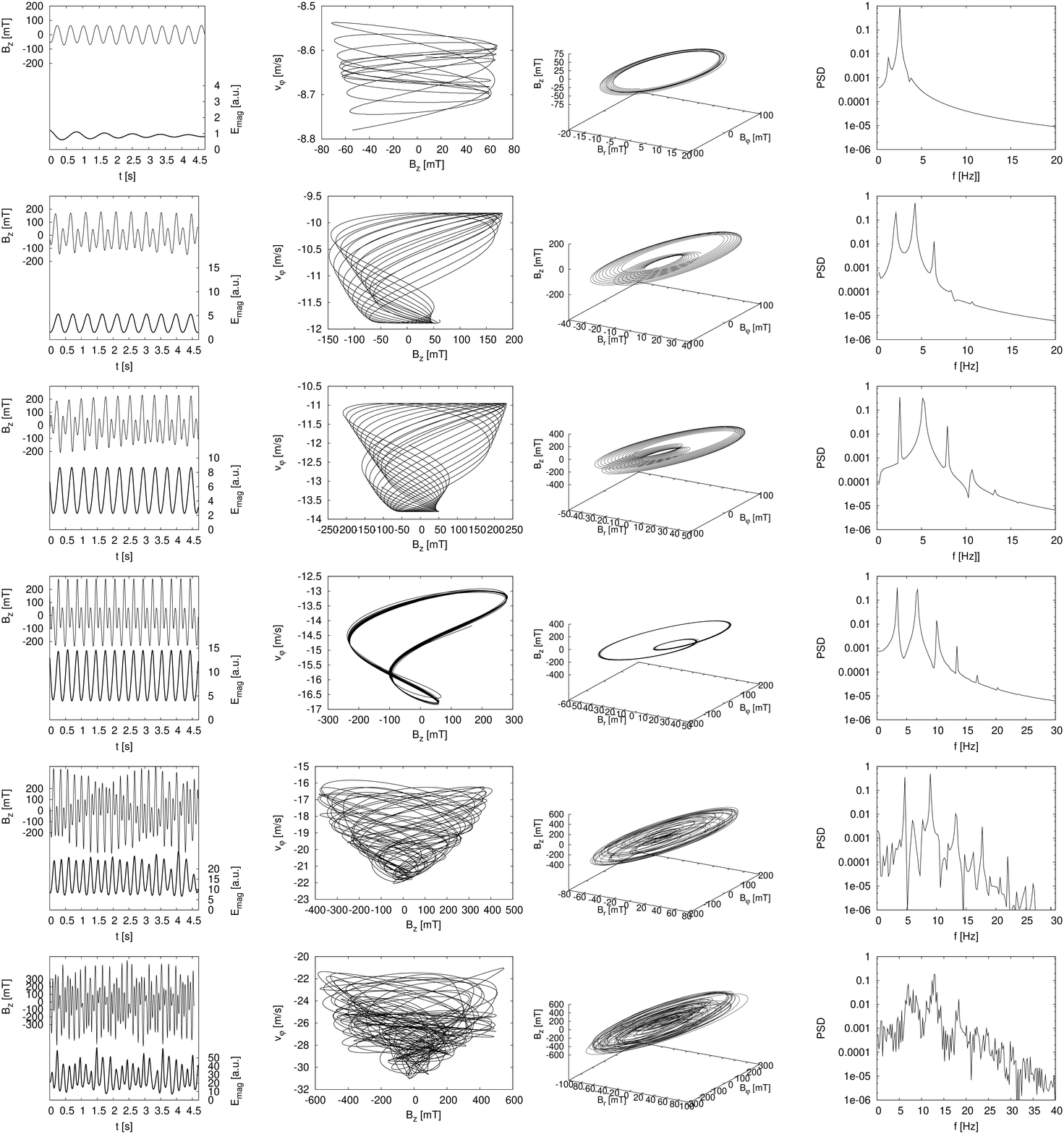}
\caption{Same as Fig. 3, but for increasing propeller rotation rate 
$P$ at fixed factor $A=1.5$. From to to bottom: $P=$ 1900, 2600, 
3000, 3600, 4500, 6000 rpm.}
\label{label5}
\end{figure*}

As for the saturation regime, our code tries to take into account 
the most important
back-reaction effect within a simple one-dimensional model
(see Gailitis et al. 2002, 2008).
While $v_z$ can be assumed rather constant 
from top to bottom
due to mass conservation,  $v_{\varphi}$ can be easily 
braked by the
Lorentz forces without any significant pressure increase.
In the inviscid 
approximation, and considering only the $m=0$ mode of the 
Lorentz force, we end up with the  differential
equation for the Lorentz force induced perturbation  
$\delta v_{\varphi}(r,z)$
of the azimuthal velocity component:
\begin{eqnarray}
\frac{\partial}{\partial t} \delta v_{\varphi}(r,z,t)&=&-
\bar{v}_z(r,z) \frac{\partial}{\partial z} \delta v_{\varphi}(r,z,t)\\ \nonumber 
&&+       \frac{1}{\mu_0 \rho} [(\nabla \times {\bf{B}}) \times
{\bf{B}}]_{\varphi}(r,z,t)  \;\; .
\label{eq2}
\end{eqnarray} 
In contrast to the procedure described in 
(Gailitis et al. 2002, 2004, 2008) which was 
focused exclusively on a stationary saturated regime, 
here we take into 
account the time dependence of $\delta v_{\varphi}(r,z,t)$ 
explicitly. 
 
Note that Eq. 2 is solved both in the innermost channel 
where it describes
the downward braking of $v_{\varphi}$, as well as in the back-flow 
channel where it
describes the upward acceleration of $v_{\varphi}$. Both effects together 
lead to a reduction of
the differential rotation and usually 
to a deterioration of the dynamo
capability of the flow, quite in accordance with Lenz's rule.
As a technical remark, note that Eq. 2 is time-stepped 
independently in the innermost 
channel and the back-flow channel, with
time-independent profiles set on two disconnection lines. 
The first disconnection line is in the 
central channel directly behind the propeller 
where the 
$v_{\varphi}(r)$ profile is set to the water-test profile (multiplied by $A$). 
The second disconnection line is at the bottom of the 
back-flow channel behind the straightening blades 
where $v_{\varphi}(r)$ is always set to zero.

The main result of this code, the downward accumulating
braking of $v_{\varphi}$, has been shown to be in 
reasonable agreement with experimental data, and with 
the results of a much 
more elaborated T-RANS simulation (see Kenjeres 
et al. 2006, 2007), apart from 
the fact that the latter simulation
results in an additional slight radial redistribution 
of $v_z$ that cannot be covered by our 
simple back-reaction model. 

In the following we will use the physical units of the 
Riga dynamo experiment, i.e. the real geometry, velocity and
conductivity with the only compromise that we will 
always use the conductivity $\sigma=8.56 \times 10^6$ S/m 
at the  reference temperature $T=157^{\circ}$C (which 
was used in 
all previous publications on the Riga dynamo). 
If we would dare, however, to lower the temperature 
to $T=102^{\circ}$C (i.e. 5$^{\circ}$C  above melting temperature) 
we would get $\sigma=10.28 \times 10^6$ S/m, and the presently
achievable propeller rotation rate of
$P=2500$ rpm  would than be equivalent (in terms of the magnetic Reynolds
number) to 3000 rpm.

With the reference conductivity $\sigma=8.56 \times 10^6$ S/m  
we obtain a diffusion time $t_{\rm diff}=\sigma \mu_0 R^2_1=0.168$ sec
if the  innermost radius $R_1=0.125$ is used, or 1.72 sec if we 
take the outer radius $R_3=0.4$ m.
The integration time for our runs was set to
6.25 sec, which in some cases is not completely sufficient for
all relaxations to be finalized.

\section{Numerical results}

In this section we present the numerical results for two
types of parameter variations. The first variation 
starts from a state that corresponds approximately 
to the maximum achievable $Rm$ number of the present 
Riga dynamo set-up, from where
we artificially scale up $v_{\varphi}$ 
by a factor $A$, 
keeping $v_z$ fixed. In the second 
variation we scale-up 
the  magnetic Reynolds number (in terms of the propeller rotation
rate $P$) 
while keeping the ratio of $v_{\varphi}$ and $v_z$
fixed at $A=1.5$.

\begin{figure}
\includegraphics[width=80mm]{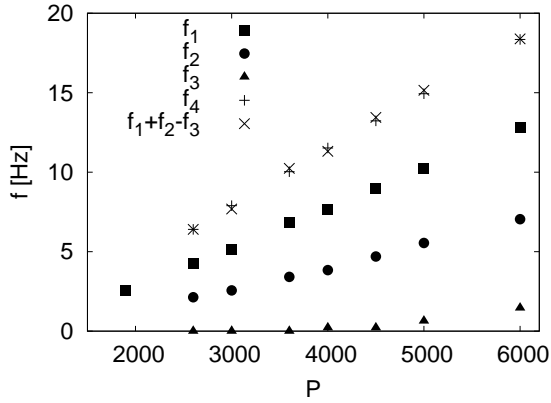}
\caption{Same as Fig. 4, but for increasing P at fixed $A=1.5$.}
\label{label6}
\end{figure}

The variation documented in Fig. 3 starts from a given 
propeller rotation rate $P=3000$ rpm with the present
ratio of $v_{\varphi}$ to $v_z$ (corresponding to $A=1$). 
Keeping the profile of $v_z$ fixed, 
we artificially increase then
$v_{\varphi}$. The rows from top to bottom in Fig. 3
correspond to increasing factors $A=$1.0, 1.3, 1.7, 2.0, 2.5, 3.0.
For each row, the first column shows the time evolution of
the magnetic field component 
$B_z$ at radius $r=0.086$ m and axial position $z=-0.39$ m 
(with respect to the mid-height of the dynamo) 
and
of the total magnetic field energy. The second column 
depicts the two-dimensional phase
portrait of $B_z$  and $v_{\varphi}$ at the same position 
$(r=0.086 \mbox{ m},z=-0.39 \mbox{ m})$. 
The third column shows a three-dimensional 
phase portrait in the
three magnetic field components $B_r$, $B_{\varphi}$, and $B_z$
at the same position. 
Finally, the fourth column 
gives  the power spectral density of  
$B_z(r=0.086 \mbox{ m},z=-0.39 \mbox{ m})$

For $A=1.0$ we obtain the well-known single-frequency 
oscillation (resulting from the 
 rotating magnetic eigenfield) with a
frequency $f_1=2.13$ Hz, which translates into 
a (more or less) horizontal line in the $B_z$-$v_{\varphi}$ 
plot and a circle in the
$B_r$-$B_{\varphi}$-$B_z$ phase portrait. 
For $A=1.3$, a second frequency $f_2=1.71$ Hz comes into play, 
which is a signature
of an arising magneto-inertial oscillation. 
Accordingly, the three-dimensional phase portrait 
shows a torus structure.
Around $A=2.0$ 
a third
frequency arises.  For $A=2.5$ and $A=3.0$ we observe then a 
irregular, probably chaotic behaviour, which 
manifests itself in the
phase portraits and the smeared out PSD.

In Fig. 4 we have compiled the three lowest peaks 
$f_1$, $f_2$, and 
$f_3$ of the
frequency spectrum.  We have also indicated a fourth 
frequency, $f_4$,
which seems to agree with the combination
$f_1+f_2-f_3$. 
It is important to note that these spectral 
properties have still
a preliminary character because the simulation time
comprises only a few oscillations. 
Anyway, there is some evidence that the appearance
of the third frequency ''rings in'' the on-set of
irregular behaviour by destabilizing the two-torus
made up by the first two frequencies $f_1$ and $f_2$.

A second parameter variation, for which the value $A$ is now fixed to 1.5, 
while the 
propeller rotation rate $P$ is increased, is illustrated in Figs. 5 and 6.
The rows from top to bottom in Fig. 4 correspond now to the axial
propeller rotation rates 
$P=$1900, 2600, 3000, 3600, 4500, 5000, 6000 rpm.

The main features are quite similar as in the former sequence, 
apart from the fact that the ratio $f_1/f_2\sim 2$ stays rather
constant within the wide range of $P=1900...4000$ rpm, a fact that 
points to a phenomenon called "mode locking". 
This is best visible in the appearance of a rather sharp
ratio $f_1/f_2=2.0$ for $P=3600$ rpm (fourth row) in Fig. 5).

Figures 7 and 8 illustrate the character of the magneto-inertial wave 
for $A=1.5$ and two selected values $P=$ 3600 and  6000 rpm, respectively.
In each case, the upper panel shows $v_{\varphi}(r=0.086 \mbox{ m},z,t)$,
and the lower panel shows $B_r(r=0.086 \mbox{ m},z,t)$.
While the case $P=3600$ rpm shows a frequency-locked wave,
the case $P=6000$ is already much more irregular.

\begin{figure}
\includegraphics[width=80mm]{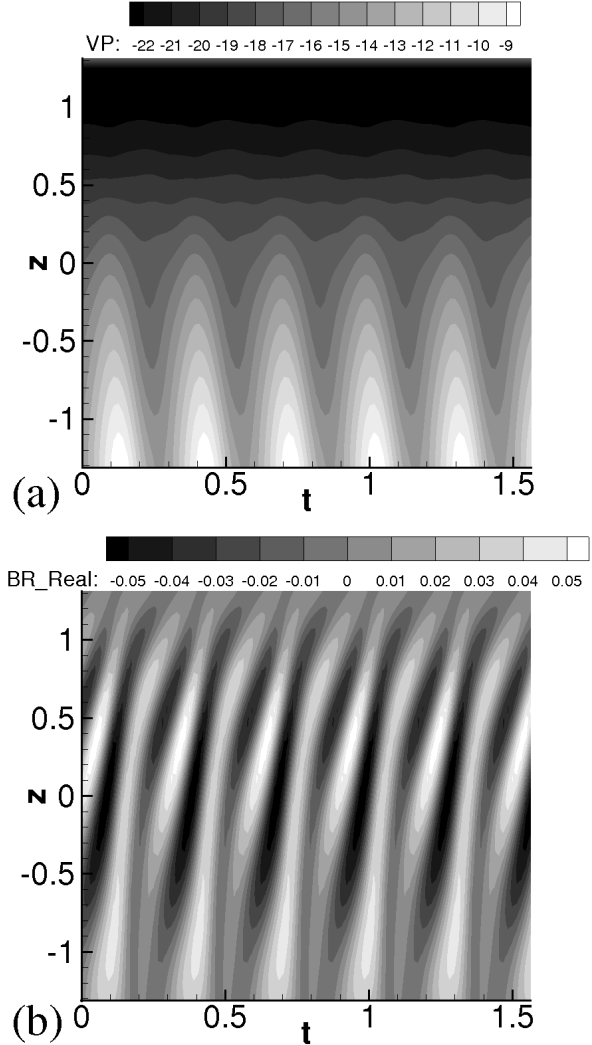}
\caption{$t-z$-dependence of $v_{\varphi}(r=0.086 \mbox{ m},z,t)$ (a) and
$B_r(r=0.086 \mbox{ m},z,t)$ (b) for $A=1.5$ and $P=3600$ rpm. The unit
of $v_{\varphi}$ is m/s, the unit of $B_r$ is Tesla.}
\label{label7}
\end{figure}

\begin{figure}
\includegraphics[width=80mm]{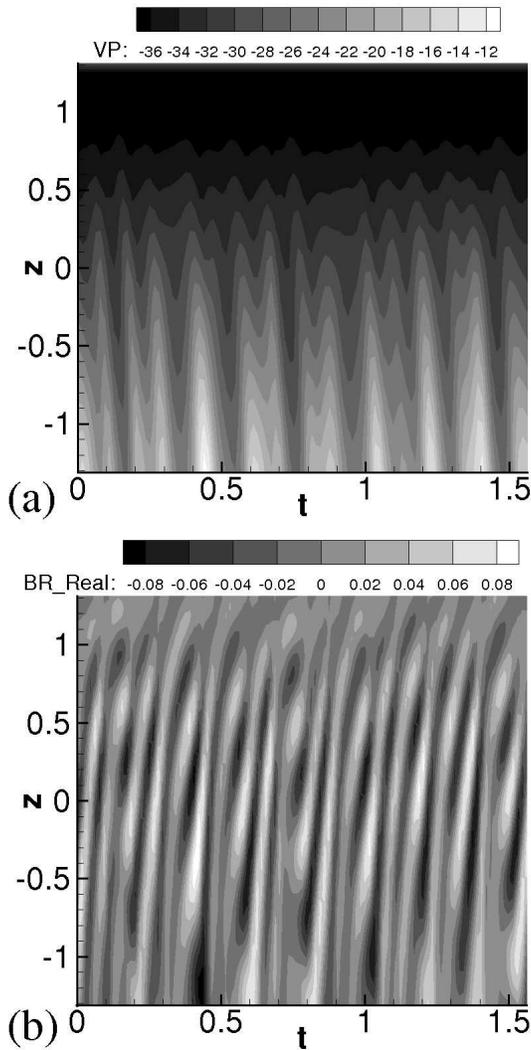}
\caption{Same as Fig. 7, but for $A=1.5$ and $P=6000$ rpm.}
\label{label8}
\end{figure}

\section{Conclusions}
Presently, the maximum 
$Rm$ in the Riga dynamo 
experiment can be achieved with a propeller rotation rate
of 2500 rpm and a temperature of around 102$^{\circ}$C, 
which is roughly equivalent to a
propeller rotation rate of 3000 rpm at a reference 
temperature of 157$^{\circ}$C.
Starting from this realistic set-up that leads to a 
single-frequency saturated state, we have studied
hypothetical modifications of the sodium flow in 
which the present ratio
of $v_{\varphi}$ and $v_z$ is scaled-up by a factor 
$A$. Beginning approximately at $A=1.3$ we observe
the appearance of a magnetic field energy oscillation 
with a frequency $f_2$ that is in the order of 
one half of the eigenfrequency $f_1$ of the saturated 
dynamo.

We hypothesize that this new oscillation with frequency $f_2$
is basically a 
magneto-inertial wave 
that traces back to the fact that the steady 
equilibrium between Lorentz and inertial
forces (that is responsible for the saturation)
becomes unstable if the ratio of $v_{\varphi}$ and $v_z$
velocity is larger than the optimum value.
This allows the system to oscillate between 
a weak field and a strong field limit.

For increasing factors $A$ we observe further
the appearance of a third
frequency $f_3$ and then the transition to a 
rather irregular, 
probably a chaotic state.
Similar transitions are observed when increasing the
propeller rotation rate $P$ while holding the factor $A$
fixed at a value of 1.5.
An interesting feature of this latter sequence is the 
appearance of a sort of frequency locking 
(characterized $f_1/f_2=2$ ) for a rather wide 
range of $2600<P<3600$.

The observed behaviour resembles the
Ruelle-Takens-Newhouse scenario of the transition to chaos 
(see Eckmann 1981). We
have to point out, though, 
the preliminary character of our study, both with respect
to limited integration time as well as to the fact
that a detailed study of the space and time resolution
is still missing.

What remains to be checked further is the
possible existence of a sub-critical Hopf-bifurcation. 

The delineated mechanism to obtain
energy oscillations in the saturated state, and even 
the sketched  route to chaos,  
seems to be in the range of technical feasibility
if we assume approximately a doubling of the motor power 
and an increase of the ratio of azimuthal to axial velocity
by means of an 
appropriate propeller and guiding blade design.

\acknowledgements

We thank the Deutsche Forschungsgemeinschaft (DFG) for 
support under grant number 
STE 991/1-1 and SFB 609, and the European Commission 
for support under grant number 028679 (MAGFLOTOM).


\newpage
  
\begin{thebibliography}{99}
  
\bibitem{} Eckmann, J.-P.: 1981, Rev. Mod. Phys. 53, 643
\bibitem{} Gailitis, A., Freibergs, Ya.: 1976, Magnetohydrodynamics 12, 127
\bibitem{} Gailitis, A., Freibergs, Ya.: 1980, Magnetohydrodynamics 16, 116
\bibitem{} Gailitis, A., Lielausis, O., Dement'ev, S., et al.: 
             2000, Phys. Rev. Lett. 84, 4365
\bibitem{} Gailitis, A., Lielausis, O., Platacis, O., et al.: 
             2001, Phys. Rev. Lett. 86, 3024
\bibitem{} Gailitis, A., Lielausis, O., Platacis, E., 
          Gerbeth, G, Stefani, F.: 2002, Magnetohydrodynamics 38, 15
\bibitem{}   Gailitis, A., Lielausis, O., Platacis, E., 
          Gerbeth, G, Stefani, F.: 2003, Surv. Geophys. 24, 247     
\bibitem{} Gailitis, A., Lielausis, O., Platacis, E., 
          Gerbeth, G, Stefani, F.: 2004, Phys. Plasmas 11, 2838
\bibitem{} Gailitis, A., Gerbeth, G., Gundrum, T., Lielausis, O.,
           Platacis, E., Stefani, F.: 2008, Compt. Rend. Phys. 9, 721
\bibitem{} S. Kenjere\v{s}, S., Hanjali\'{c}, K., Renaudier, S. 
         Stefani, F., Gerbeth, G., Gailitis, A.: 2006, 
       Phys. Plasmas 13, 122308. 
\bibitem{} Kenjere\v{s}, S., Hanjali\'{c}, K.: 2007, 
Phys. Rev Lett. 98,104501. 
\bibitem{} Ponomaranko, Yu.B.: 1973, J. Appl. Mech. Tech. Phys. 14, 775
\bibitem{} Ravelet, F., Berhanu, M., Monchaux, R., et al.: 2008, Phys. 
Rev. Lett. 101, 074501
\bibitem{} Rohde, R., R\"udiger, R., Elstner, D.: 1999, Astron. Astrophys. 347, 
860
\bibitem{} Reuter, K., Jenko, F., Forest, C.B.: 2009, New J. Phys. 11, 013027
\bibitem{} Schmitt, D.: 2010, Geophys. Astrophys. Fluid Dyn. 104,  135
\bibitem{} Stefani, Gailitis, A., Gerbeth, G.: 2008, ZAMM 
         88, 930	
\end{thebibliography}
\end{document}